\documentclass{amsart}

\usepackage{bbm}
\usepackage{graphicx}
\usepackage{verbatim}
\def\cbstart{ } \def\cbend{ } 

\usepackage[latin1]{inputenc}

\newtheorem{proposition}{Proposition}
\newtheorem{theorem}{Theorem}
\newtheorem{lemma}{Lemma}

\newcommand\ind[1]{\mathbbm{1}_{\{#1\}}}

\def\cal{\mathcal}
\def\C{{\mathbb C}}

\def\R{{\mathbb R}}

\def\P{{\mathbb P}}
\def\E{{\mathbb E}}
\def\Var{\mathrm{Var}}

\def\eps{\varepsilon}

\def\etal{{\em et al.}}
\def\T{\mathcal{T}}
\title[Analysis of  Steiner subtrees]{Analysis of  Steiner subtrees of Random Trees for Traceroute Algorithms}

\author{Fabrice  Guillemin}
\address[F.~Guillemin]{Orange Labs, 2, Avenue Pierre Marzin, F-22300 Lannion}
\email{Fabrice.Guillemin@orange-ftgroup.com}
\author{Philippe Robert}
\address[Ph.~Robert]{INRIA-Rocquencourt,  RAP project, Domaine de Voluceau, 78153 Le Chesnay, France}
\email{Philippe.Robert@inria.fr}
\urladdr{http://www-rocq.inria.fr/\~{}robert}
\date{\today}

\begin{document}

\begin{abstract}
We consider  in this  paper the problem  of discovering,  via a traceroute  algorithm, the
topology of a network, whose graph is  spanned by an infinite branching process.  A subset
of nodes  is selected  according to  some criterion.  As  a measure  of efficiency  of the
algorithm,  the Steiner distance  of the  selected nodes,  i.e. the  size of  the spanning
sub-tree of  these nodes, is investigated.  For  the selection of nodes,  two criteria are
considered: A node is randomly selected with a probability, which is either independent of
the depth of the node (uniform model)  or else in the depth biased model, is exponentially
decaying with  respect to  its depth.  The  limiting behavior  the size of  the discovered
subtree is investigated for both models.
\end{abstract}


\maketitle

\hrule

 \tableofcontents 

\vspace{-5mm}

\hrule

\vspace{5mm}

\section{Introduction}
In  the past  ten  years, the  Internet has  known  an extraordinary  expansion and  still
experiences  today a  sustained  growth.  The  counterpart  of this  success  is that  the
different  autonomous  systems  composing  the  global Internet  have  been  independently
developed  by different  operators.   This raises  some  issue since  the  Internet is  by
construction a flat network, where the different components are interdependent in terms of
connectivity availability, security,  quality of service etc.  It thus  turns out that the
knowledge of the physical layout of a  network is of prime interest for network operators. The
physical topology of a component of the Internet is in general very difficult to describe.
To  establish a  representation of  the whole  or a  part of  the Internet,  some topology
exploration methods have to be  devised.  Various topology discovery experiments have been
initiated  by  different organizations  in  order  to infer  the  topology  of the  global
Internet,   notably    the   Skitter   project   by   CAIDA    \cite{caida},   the   DIMES
project~\cite{Dimes}  and  many other  initiatives.   The  method  generally proposed  for
analyzing  the topology  of  a network  is based  on  the traceroute  facility offered  by
routers.  Roughly speaking,  a  traceroute  procedure  consists of  sending
traceroute messages between hosts as follows:

\vspace{4mm}

\hrule

\vspace{3mm}

\noindent
{\sc Traceroute Algorithm} \\
If $H$ and $G$ are  hosts participating in the topology discovery experiment, 
$H$ sends to $G$ a traceroute message so that all the hosts/routers on the path 
$(H,G)$ are identified.

\vspace{3mm}

\hrule

\vspace{4mm}

The purpose  of this paper is to  investigate the efficiency of  the traceroute algorithm.
While a large  number of experimental papers are available in  the technical literature on
the  analysis of  the topology  of the  Internet, a  very few  studies  provide analytical
insight  into  the  efficiency  of   these  topology  discovery  methods;  see  Vespignani
\etal~\cite{Vespignani} for a discussion and Azzana \etal~\cite{Azzana:01} for an analysis
in the case of specific deterministic trees.

In this paper, a more realistic model is proposed to include some randomness in the degree
of  the nodes  of  the graph  representing the  topology  of a  network. One  specifically
considers a network  with a random tree architecture spanned  by a Galton-Watson branching
process.  \cbstart{We shall restrict the analysis to the case of offspring distributions, which have a finite second momenmt. This notably precludes the case of power law distributions with infinite second moments, typically distributions $G$ such that $\P(G \geq n) \sim C n^{-\alpha} $ with $\alpha \in (0,2)$.}\cbend

The Internet graph is definitely not a tree, since many studies (see the Skitter
project)  show that  there  is a  core  of highly  connected  routers. Nevertheless,  some
components of the Internet  have a  topology close to a tree structure. This is notably
the case  of access or collect  networks, which play  the role of capillarity  networks in
charge  of collecting  and distributing  traffic  between customers  and the  core of  the
Internet. This  latter component is not  critical for the  problem we study in  this paper
since core routers are easy to discover by traceroute procedures.  This is why we focus on
collect networks, which can be represented  by a tree architecture, spanned by a branching
process. In addition, to get more insight  into the topology discovery process in the case
of a large network, it is assumed that the underlying branching process does not terminate
with probability $1$; in particular the depth of the tree is infinite.

The discovery process is as follows: a random number of nodes are selected among the nodes
of the tree.  After the selected nodes have performed the traceroute algorithm, the set of
the nodes  discovered is  the spanning  tree of the selected  nodes. The performance
criterion used in  this paper is simply the  size of this sub-tree. In graph  theory it is
known as the  {\em Steiner distance} of the  selected nodes (with the  slight difference that
the selected  nodes are not  counted).  It has  been the subject  of a recent  interest by
Mahmoud  and  Neininger~\cite{Mahmoud2}   and  Christophi  and  Mahmoud~\cite{ChMa}  which
considered the asymptotic  behavior of the distance between two random  nodes of the tree.
Panholzer~\cite{Panholzer}, Panholzer and Prodinger~\cite{Panholzer2} proved central limit
theorems  when multiple  points are  considered.   The asymptotics  investigated in  these
papers concern the  size of the random tree.  In our paper, we will  study two situations:
when the size  of the tree and the number  of selected nodes go to  infinity and also when
the   infinite  tree   is  fixed   and   the  number   of  selected   nodes  grows.    See
Panholzer~\cite{Panholzer} for a thorough discussion of the literature in this domain.

Two stochastic models for selecting the nodes  in the network are considered. In the first
model,  the uniform model,  we adopt  the point  of view  of an  external observer  to the
network; a set of  nodes is chosen at random and a  traceroute algorithm is performed.  In
the second model,  the depth biased model, the  observer is located at one  node (the root
node) and  it chooses more  likely nodes  not too far  away.  As it  will be seen,  in the
uniform model , the selected nodes are  basically in the ``bottom'' of the tree where most
of the nodes  are, while in the second  model they are more concentrated at  the ``top'' of
the tree.

In the first model, referred to as the uniform model, nodes whose depth is less than $N>0$
are   randomly  chosen  with   probability  $1-\exp(-\lambda)$   for  some   $\lambda  >0$
independently of the position of the node  in the tree.  The quantity analyzed here is the
ratio $\rho_N(\lambda)$ of the mean size $\E(R_N)$ of the sub-tree discovered and the mean
number  $\E(T_N)$ of  nodes  of the  tree  whose depth  is less  than  $N$.  The  quantity
$\rho_N(\lambda)$ denotes the fraction of  the tree discovered.  The asymptotic results of
this paper first determine the limit  $\rho(\lambda)$ of $\rho_N(\lambda)$ as $N$ tends to
infinity.  In a  second step, the asymptotic behavior of  $\rho(\lambda)$ for $\lambda \to
0$  is investigated.   This  last  point gives  an  indication of  the  efficiency of  the
algorithm when only a few nodes are selected in the topology discovery experiment.

For the  uniform model, it is shown  in Theorem~\ref{rho1} that, for  small $\lambda$, the
exploration rate $\rho(\lambda)/\lambda$ is equivalent to $\log_m\lambda$ where $m$ is the
mean  value of  the offspring  distribution of  a node,  so that  at the  first  order the
algorithm is very efficient.   A second order analysis, Proposition~\ref{vartree}, reveals
that the standard deviation  of the size of the discovered tree  scales with the mean size
of the tree, except when the  offspring distribution is deterministic. This latter case is
degenerate in  the sense that  the standard deviation  is negligible when compared  to the
mean value.

In the second model, referred to as the depth biased model, the probability of selecting a
node depends on its  depth in the tree so that the mean number  of selected nodes at depth
$n$ is $\alpha^n$ for some $\alpha >0$.  It is shown in Theorem~\ref{oscil} that the ratio
of the average of the size  $R(\alpha)$ of the sub-tree discovered and the
average number  of selected nodes is equivalent to  $1/(1-\alpha)$.

The  paper is  organized  as follows:  In  Section~\ref{formulation}, the  models for  the
selection of the nodes  of the tree are introduced.  The uniform  model is investigated in
Section~\ref{uniform}  and  the depth  biased  model  in  Section~\ref{biased}.  The  main
ingredients for the analysis of these models are Kesten-Stigum Theorem and some results on
the rates of convergence for Galton-Watson branching processes and a general limit theorem
proved in Section~\ref{Convsec}.

\subsection*{Acknowledgments}
The authors wish to thank two anonymous referees for their work,  their detailed comments 
have helped us a lot to improve and correct mistakes in the  first version of the paper.

\section{Problem Formulation}\label{formulation}
Throughout this  paper, we consider  a Galton-Watson branching  process, whose graph  is a
tree denoted by ${\cal T}$. Each element of the $n$th generation (or $n$th level) gives birth
to $G$ nodes at the $(n+1)$th generation  independently of the other elements of the $n$th
level, where the offspring $G$ is some {\em integrable} random variable.  (See Athreya and
Ney~\cite{Athreya:04} and  Lyons and Peres~\cite{Lyons:07}  for an introduction  to random
trees.)

 It is  assumed  that $\P(G{=}0){=}0$ and $P(G\geq 2)>0$, in particular the  tree  is supercritical, i.e. $m=\E(G)>1$.
For $n\geq 0$, the variable $Z_n$ denotes  the number of nodes at level $n$, in particular
$Z_0=1$.  For  $1\leq\ell\leq Z_n$,  a node  of the tree  can be  represented as  a pair
$(n,\ell)$, where  $n$ is its  generation and $\ell$  its rank within the  generation. (For notational conventions, see
Neveu~\cite{Neveu:20}  for example.)  Let  ${\cal T}_{k}^{n,\ell}$  denote  the sub-tree  of
${\cal T}$ with depth less than or equal to  $k$ and with root at node $(n,\ell)$. The size of
${\cal T}_{k}^{n,\ell}$ is denoted by $T_{k}^{n,\ell}$.  When $(n,\ell)$ is the root node,
i.e. $(n,\ell)=(0,1)$, the  upper index $(0,1)$ is omitted.   With the above notation,
one gets easily that for all $N>1$ and $n= 1,\dots, N$
\begin{equation}\label{eq1st}
T_N=\sum_{i=0}^{n-1}Z_i+ \sum_{\ell=1}^{Z_{n}} T_{N-n}^{n,\ell}.
\end{equation}

\begin{figure}[ht]
\begin{center}
\scalebox{.5}{\includegraphics{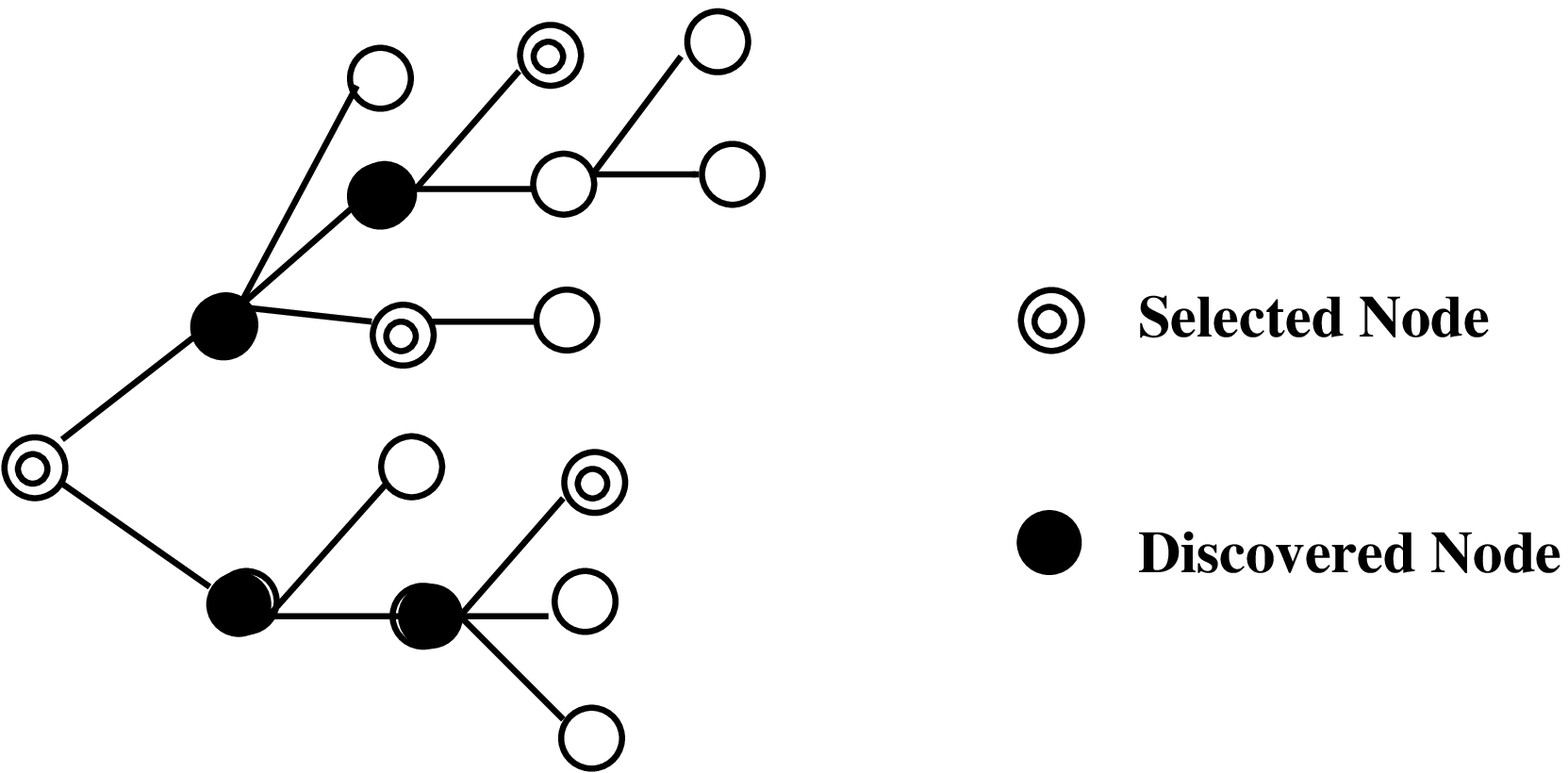}}
\end{center}
\caption{Traceroute Algorithm.}\label{trafig}
\end{figure}

Let us consider a counting measure $\mathcal{N}$ on the tree representing the distribution
of the  points selected in  the tree: For a  subset $A$ of  the nodes of the  tree, ${\cal
N}(A)$ denotes  the total  number of  points in $A$.  By selecting  nodes, a  sub-tree from
$\mathcal{T}$ is obtained  through the traceroute algorithm; this sub-tree  is  referred to as sampled
tree. See Figure~\ref{trafig}.

To complete  the description of the  problem, it remains to  specify how the  nodes of the
original tree are selected.  In the following, we shall consider two selection criteria:
\begin{description}
\item[Uniform model] Nodes are  chosen at random on all the nodes  of the tree whose depth
is  less than  or equal  to $N$,  $N$  being a  fixed integer.   A node  is selected  with
probability $1-\exp(-\lambda)$ independently of his depth  in the tree. The mean number of
nodes involved in the discovery experiment is then $(1-\exp(-\lambda)) (m^{N+1}-1)/(m-1)$.
(Recall that the mean size of the  $n$th generation is $m^n$, $n\geq 0$, where $m =\E(G)$,
the mean of the offspring variable $G$.)

To investigate the topology discovery process, we shall consider for a fixed
$N>0$ the  $N$ first levels  of the  original tree $\mathcal{T}$  and count the  number of
nodes which are discovered, given by
\begin{equation}\label{relfond}
R_N=\sum_{n=0}^N \sum_{\ell=1}^{Z_{N-n}}\ind{{\cal N}(\T_{n}^{N-n,\ell})\not=0}.
\end{equation}
In the following, we shall be particularly interested in the quantity
\begin{equation}
\label{defrhoN}
\rho_N(\lambda)= \frac{\E(R_N)}{\E(T_N)},
\end{equation}
i.e., the ratio of the mean number of  discovered nodes to the mean number of nodes in the
tree,  when the analysis  is restricted  to the  $N$ first  levels of  the tree.  Then the
behavior of this ratio when the number $N$ of levels tends to infinity is investigated.

\item[Depth biased model] Nodes at given level $n$ are selected with probability
$1-\exp[-(\alpha/m)^n]$ for some $\alpha\in(0,1)$.  The mean number of nodes selected at
level $n$ is $m^n(1-\exp[-(\alpha/m)^n]) \sim \alpha^n$ and therefore is 
exponentially decreasing with respect to the depth. The rational behind that is the fact
that, for this model, the traceroute procedure will rarely  select nodes ``far away'' from the root
node, in contrary to the uniform case where geometric aspects are completely ignored for the
selections of the hosts. 

By denoting by $R(\alpha)$ the total number
of nodes discovered, the efficiency of the traceroute algorithm is measured in this case 
through the ratio  of the mean $\E(R(\alpha))$ to  the average number of
selected nodes. The limiting behavior when the  average number of selected  nodes becomes
large, i.e. when $\alpha\nearrow 1$,  is investigated. 

\end{description}
Additionally it is  assumed that the root node  of the tree is always selected;  it is not
difficult to show that for both models described above, the root node belongs to
the sample tree  with a very high probability and then the above assumption is not really restrictive.  This implies that a node $(n,\ell)$ of
the tree  ${\cal T}$ at  level $n$  belongs to the  sampled tree whenever  ${\cal N}({\cal
  T}_{N-n}^{n,\ell})$ is not  $0$.  In other words, a node of  the original tree belongs
to the discovered tree if at least one of his descendants has been selected. \cbstart In the following, we shall use the following notation: for a subtree $\T^{N-n,\ell}_{n}$ rooted at a vertex $(N-n,\ell)$ of the $(N-n)$th generation of the tree $\T$ and with depth $n$, the quantity $\P(\mathcal{N}(\T^{N-n,\ell}_n)\neq 0)$ is the probability that a least one vertex of the subtree $\T^{N-n,\ell}_{n}$ is marked and $(N-n, \ell)\in \T$.\cbend

Before proceeding to the analysis of the  topology discovery process, we prove in the next
section  a technical  result, which  is important  in  the analysis  of the  speed of  the
exploration process.

\section{A Convergence Result}\label{Convsec}
To prove asymptotic expansions in the following sections, the
following proposition will repeatedly be used. Its proof is based on integral
representations and Fubini's Theorem instead of complex analysis techniques as it is
usually the case in the context of harmonic series.  See Robert~\cite{Robert:09} for a presentation of
these methods.  
\begin{proposition}\label{asympprop}
Let $V$  be a positive  random variable with  $\E(V^2)<+\infty$ and $h$ be  a non-negative
twice  differentiable function  on $\R_+$  such that  $h(0)=0$. In  addition,  it is
  assumed that  the function  $h'$ is integrable  with $h'(0)\not=0$  and \cbstart that there exists some constant $K>0$ such that $|h''(x)|<K$  for all $x\in [0,\infty)$. \cbend

The function $\Psi(h)(x)$ defined by
\begin{equation}\label{Branaire}
\Psi(h)(x)=\sum_{n=0}^{+\infty} \frac{1}{m^n}\E\left(h\left(x Vm^n\right)\right),\quad x\geq 0,
\end{equation}
is such that 
\[
\lim_{x\to 0} \frac{\Psi(h)(x)}{x\log_m(1/x)} = \E(V)h'(0).
\]
\end{proposition}
\begin{proof}
Since $h$ is non-negative and $|h'|$ integrable with respect to Lebesgue measure on $\R_+$,
Fubini's Theorem applied twice  shows that $\Psi(h)$ can be expressed as 
\begin{align}
\Psi(h)(x)&=\sum_{n=0}^{+\infty} \frac{1}{m^n}\E\left(h\left(x Vm^n\right)\right)=
\E\left(\sum_{n=0}^{+\infty} \frac{1}{m^n} h\left(x Vm^n\right)\right)\notag\\
&= \E\left(\sum_{n=0}^{+\infty} \frac{1}{m^n} \int_0^{+\infty} h'(u)\ind{u\leq x
  Vm^n}\,du \right)\notag \\
&= \E\left(  \int_0^{+\infty} h'(u)\sum_{n=0}^{+\infty} \frac{1}{m^n} \ind{u\leq x
  Vm^n}\,du \right).\label{Lang}
\end{align}
The function $\Psi(h)$ is thus well defined. 

Since $h'(0)>0$,  Fatou's Lemma applied successively gives the relation
\begin{multline*}
\liminf_{x\to 0} \frac{\Psi(h)(x)}{x}\geq 
\sum_{n=0}^{+\infty} \liminf_{x\to 0} \frac{m-1}{m^n}\E\left(\frac{h\left(x
  Vm^n\right)}{x}\right)\\
\geq \sum_{n=0}^{+\infty} \frac{m-1}{m^n}\E\left(\liminf_{x\to 0}\frac{h\left(x
  Vm^n\right)}{x}\right)
=\sum_{n=0}^{+\infty} (m-1)\E\left(V\right)h'(0)=+\infty,
\end{multline*}
therefore the ratio $\Psi(h)(x)/x$ diverges as $x\to 0$. 

By using representation~\eqref{Lang} of $\Psi(h)$, we have
\begin{multline*}
(m-1)\Psi(h)(x)
= m\E\left(\int_{0}^{xV} h'(u)\,du \right)\\+\E\left(\int_{xV}^{V}
\frac{1}{m^{\lfloor \log_m(u/xV)\rfloor}}h'(u)\,du \right)+ \E\left(\int_{V}^{+\infty}
\frac{1}{m^{\lfloor \log_m(u/xV)\rfloor}}h'(u)\,du \right),
\end{multline*}
where $\lfloor y \rfloor$ is the integer part of $y\in\R$. One first shows that only the
central term of the right hand side plays a role in the asymptotic behavior of $\Psi(h)$
at the first order. 

For the first term, note that, if $\|h''\|_{\infty}$ is the $L_\infty$ norm of $h''$,
\begin{multline*}
\left|\frac{1}{x}\E\left(\int_{0}^{xV} h'(u)\,du \right)\right|
\\\leq \frac{1}{x}\E\left(\int_{0}^{xV} \left(h'(0)+u \|h''\|_{\infty}\right)\,du \right)
\leq h'(0)\E(V)+\frac{x}{2}\E(V^2)\|h''\|_{\infty}
\end{multline*}

For $u\geq V$, one has 
\[
x m^{\lfloor \log_m(u/xV)\rfloor}\geq x m^{\lfloor \log_m(1/x)\rfloor}\geq 
x m^{\log_m(1/x)-1}=m^{-1},
\]
and hence,
\[
\frac{1}{x}\E\left(\int_{V}^{+\infty} 
\frac{1}{ m^{\lfloor \log_m(u/xV)\rfloor}}|h'(u)|\,du \right)
\leq m \int_0^{+\infty} |h'(u)|\, du.
\]
By gathering these estimations, it follows that the following equivalence 
\begin{multline*}
\frac{\Psi(h)(x)}{x}\sim \E\left(\int_{xV}^{V}  \frac{1}{x m^{\lfloor \log_m(u/xV)\rfloor}}h'(u)\,du \right) 
\\ = \E\left(V\int_{xV}^{V} m^{\{ \log_m(u/xV)\}}\frac{h'(u)}{u}\,du\right),
\end{multline*}
holds as $x\to 0$, 
with $\{y\}=y-\lfloor y \rfloor$, the fractional value of $y\in\R$. The above equivalence can be rewritten as
\begin{multline*}
\frac{\Psi(h)(x)}{x}\sim\E\left(V\int_{xV}^{V} m^{\{ \log_m(u/xV)\}}\frac{h'(u)-h'(0)}{u}\,du\right)
\\ +h'(0)\E\left(V\int_{xV}^{V} \frac{m^{\{ \log_m(u/xV)\}}}{u}\,du\right)
\end{multline*}
Due to  the boundedness of  $h''$ and the  integrability of $V^2$,  the first term  in  the
right hand side of the above equation is  bounded as $x$ goes to $0$. Hence, only the second term
has to be considered. For $x<1$, we have
\begin{align*}
\int_{xV}^{V} &\frac{m^{\{ \log_m(u/xV)\}}}{u}\,du
=\int_1^{1/x} \frac{m^{\{\log_m(u)\}}}{u}\,du\\
&=\sum_{{k\geq 0:m^k\leq 1/x}}
\int_{m^k}^{m^{k+1}} \frac{m^{\{\log_m(u)\}}}{u}\,du+O(1)
=(m-1)\lfloor -\log_m(x)\rfloor +O(1)
\end{align*}
and the result follows.
\end{proof}

Asymptotic behavior of algorithms with an underlying   tree   structure has been
extensively investigated,   see   Flajolet  \etal~\cite{Flajolet:14},   Mohamed   and
Robert~\cite{Mohamed:01}  and Mahmoud~\cite{Mahmoud:02}  for a  general  presentation.  
By using the terminology of Flajolet \etal~\cite{Flajolet:14}, for non-negative sequences
$(\lambda_n)$ and $(\mu_n)$, a series like
\begin{equation}\label{eqaus}
G(x)=\sum_{n\geq 0} \lambda_n g(\mu_n x),
\end{equation}
for some function $g$ is defined as an {\em harmonic sum}. Because of the integration of the
random variable $V$ and given that one wants the weakest assumptions on this random variable,
series~\eqref{Branaire} could be seen as a \cbstart special case \cbend  of harmonic sums. 
The fact that the sequences $(\lambda_n)$ and $(\mu_n)$ are specific in Expression~\eqref{Branaire} 
is not a real restriction, see Robert~\cite{Robert:09}.  

Flajolet \etal~\cite{Flajolet:14} derives the asymptotic expansion of $G(x)$  when
$x$ goes to $0$ or $+\infty$ by using Mellin transform techniques. For $s\in\C$, if
$h^*(s)$ is the Mellin transform of $h$, i.e. for $s$ in some vertical strip of $\C$,
\[
h^*(s)=\int_0^{+\infty} h(x) \,x^{s-1}\,dx,
\]
it is easy to check that the Mellin transform of $\Psi(h)$ is given by
\[
\Psi(h)^*(s)=\frac{1}{1-m^{-(s+1)}}\E\left(V^{-s}\right) h^*(s).
\]
Following the methods of Flajolet \etal~\cite{Flajolet:14}, to derive  the asymptotic behavior
of $\Psi(h)(x)$ as $x$ goes to infinity, one has to identify the first singularity of
$\Psi(h)^*$ on the right of the maximal vertical strip where it is defined. In particular, 
some conditions on the finiteness of some fractional moments of the random variable $V$ have to be
assumed (as well as growth conditions on $h^*$). From this point of view, our approach is
minimal since only the finiteness of $\E(V^2)$ and differentiability  conditions on $h$ are assumed. It turns out that it is important
as it will be seen in the following sections, since in practice little is known on the
fractional moments of the corresponding variable $V$. 

\section{The Exploration Rate in the Uniform Model}\label{uniform}
In this section, nodes are selected at  random with uniform probability in
the tree  with depth  less than $N$.   The variable  $R_N$ is the  size of  the underlying
sub-tree  (or sampled tree)  containing the  selected nodes.   The asymptotic  behavior of
$\rho_N(\lambda)=\E(R_N)/\E(T_N)$,  the fraction of  discovered nodes,  when $N$  tends to
infinity   is  investigated.    In   the  second   part   of  this   section,  the   ratio
$\mathrm{var}(R_N)/\E(T_N)$ is analyzed.

\subsection{First Order Asymptotics}

In the uniform case, the limiting behavior of the ratio $\rho_N(\lambda)$ when $N$ tends to infinity is given by the following result.

\begin{theorem}\label{ThFirst}\label{rho1}
The  ratio of the average size $R_N$ of the sampled tree  to
the total average size of the tree $\E(T_N)$ satisfies the relation
\begin{equation}
\label{defrholambda}
\rho(\lambda)\stackrel{\text{def.}}{=}\lim_{N\to+\infty}\rho_N(\lambda) =\sum_{n=0}^{+\infty}    \frac{m-1}{m^{n+1}}\left(1-\E\left(\exp\left(-\lambda
  \sum_{i= 0}^{n}Z_i\right)\right)\right).
\end{equation}
If additionally the condition $\E\left(G^2\right)<+\infty$ holds then
\begin{equation}\label{rate}
\lim_{\lambda \to 0} \frac{\rho(\lambda)}{\lambda\log_m(1/\lambda)} = 1.
\end{equation}
\end{theorem}

Relation~\eqref{rate} shows that the rate of increase of the discovery process is infinite
near the origin. This implies that with only  a few selected nodes one has the impression of rapidly discovering the whole network.

\begin{proof}
By conditioning on the tree, the conditional probability that node $(N-n,\ell)$ does not
belong to the sampled tree is 
\[
\left.\P\left( \rule{0mm}{4mm}{\cal N}\left({\cal T}^{N-n,\ell}_{n}\right)\not=0\right| {\cal T}\right)=
1-\exp\left(-\lambda T_{n}^{N-n,\ell}\right).
\]
By summing-up these relations, one obtains  that the expected value of $R_N$, i.e., the average
number of nodes in the sampled tree, is given by
\begin{align*}
\E(R_N)&=\sum_{n=0}^{N} \E(Z_{N-n})
\left(1-\E\left(\exp\left(-\lambda T_{n}\right)\right)\right)\\
&=\sum_{n=0}^{N} m^{N-n}\left(1-\E\left(\exp\left(-\lambda   \sum_{i = 0}^{n}Z_i\right)\right)\right).
\end{align*}
The limit when $N \to \infty$ of the ratio $\rho_N(\lambda)$ is then given by
$$
\rho(\lambda)\stackrel{\text{def.}}{=}\lim_{N\to +\infty} \rho_N(\lambda) =\sum_{n=0}^{+\infty}
  \frac{m-1}{m^{n+1}}\left(1-\E\left(\exp\left(-\lambda  \sum_{i=0}^{n}Z_i\right)\right)\right)
$$
since $\E(T_N) \sim m^{N+1}/(m-1)$ for large $N$. This proves the first equality stated in Theorem~\ref{ThFirst}.

We now study the behavior of $\rho(\lambda)$ when $\lambda$ goes to $0$.  Since $\E(G^2)<+\infty$, 
Kesten-Stigum's Theorem ensures the existence of a
random variable $W$ such that $\P(W>0)=1$ (because of the assumption on the
distribution of $G$) and $\E(W)=1$ (See Lyons and Peres~\cite{Lyons:07}) and that, almost surely,
\begin{equation}
\label{defW}
\lim_{n\to+\infty} \frac{Z_n}{m^n}=W.
\end{equation}
 Let us define 
$$
f(\lambda)\stackrel{\text{def.}}{=}\sum_{n=0}^{+\infty}
 \frac{m-1}{m^{n+1}}\left(1-\E\left(\exp\left(-\lambda W \frac{m^{n+1} -1}{m-1}\right)\right)\right).
$$
Then,
\begin{multline}\label{aux:1}
\frac{|\rho(\lambda)-f(\lambda)|}{\lambda}  \\ \leq 
\sum_{n=0}^{+\infty}
  \frac{m-1}{\lambda m^{n+1}}\left|\E\left(\exp\left(-\lambda
  \sum_{i=0}^{n}Z_i\right)\right)-\E\left(\exp\left(-\lambda W
    \frac{m^{n+1}-1}{m-1}\right)\right)\right|.
\end{multline}

Since $W$ is integrable, Lebesgue's dominated convergence Theorem gives that
\begin{multline}\label{eqaux:1}
\lim_{\lambda\to 0} \frac{1}{\lambda}
\E\left(\exp\left(-\lambda
  \sum_{i=0}^{n}Z_i\right)-\exp\left(-\lambda W \frac{m^{n+1}-1}{m-1}\right) \right)\\=
\E\left(\sum_{i=0}^{n}Z_i- W \frac{m^{n+1}-1}{m-1}\right)=0.
\end{multline}
We have
\begin{multline*}
\frac{1}{m^{n+1}\lambda}
\left|\E\left(\exp\left(-\lambda\sum_{i=0}^{n}Z_i\right)\right)-\E\left(\exp\left(-\lambda W \frac{m^{n+1}-1}{m-1}\right)\right)\right|
\\ \leq \frac{1}{m^{n+1}} \sum_{i=0}^{n}\E |Z_i -m^i W|.
\end{multline*}
From Athreya and Ney~\cite[Theorem~1, page~54]{Athreya:04},  for $n\geq 1$, there
exists a sequence $(W^i)$ of i.i.d. random variables with the same distribution as $W$
such that
\begin{equation}\label{CV}
Z_n-m^n W=\sum_{i=1}^{Z_n} (1-W^{i}).
\end{equation}
By using Cauchy-Shwartz's Inequality, we obtain
\begin{align}
\E\left(\left|Z_n-m^n W\right|\right)|&
\leq  \sqrt{\E\left((Z_n -m^n   W)^2\right)}\notag \\
&=\Var(1-W)\sqrt{\E(Z_n)}\notag \\
&=\Var(1-W) m^{n/2}.\label{CS}
\end{align}
From the above inequality, we deduce that 
\begin{multline*}
\frac{1}{m^{n+1}\lambda}
\left|\E\left(\exp\left(-\lambda\sum_{i=0}^{n}Z_i\right)\right)-\E\left(\exp\left(-\lambda W \frac{m^{n+1}-1}{m-1}\right)\right)\right|
\\\leq \frac{\Var(1-W)}{\sqrt{m}-1} \frac{1}{m^{(n+1)/2}}.
\end{multline*}
Relation~\eqref{eqaux:1} and Lebesgue's Theorem then imply that
\[
\lim_{\lambda\to 0}\frac{\rho(\lambda)-f(\lambda)}{\lambda}=0.
\]
Hence, up to an expression which is of the order of $o(\lambda)$, the behavior at $0$ of
$\rho(\lambda)$ is equivalent to  the behavior of $f(\lambda)$ as $\lambda$ becomes  small.

By using Proposition~\ref{asympprop}, we have by taking $h(u)=1-e^{-u}$  and $V=W m/(m-1)$,
\[
\sum_{n=0}^{+\infty} \frac{m-1}{m^{n+1}}\left(1-\E\left(\exp\left(-x V m^{n}\right)\right)\right) =\Psi(h)(x)\sim x \log_m (1/x)
\]
as $x \to 0$. To conclude the proof, we note that
$$
\lim_{x \to 0} \frac{\Psi(h)(x) -f(x)}{x\log_m x}=0
$$
and the result follows.
\end{proof}

\subsection{Second Order Properties}

The results obtained in the previous section show that the size of the sampled tree is of
the same order of magnitude as the original tree when the probability of selecting a node is fixed. When this probability  is very small (i.e., for small $\lambda$), the
speed of the discovery process is even very fast. In this section, we evaluate the second
moment of the random variable $R_N$ in order to estimate the dispersion of the size of the
sampled tree around the mean value. 

\cbstart In the rest of this section, we use the following notation: If  $(n,\ell)$  and  $(n',\ell')$  are  two  nodes of  the
tree, the relation  $(n',\ell') < (n,\ell)$ indicates that  the nodes are distinct and
that $(n',\ell')$ is a node of the sub-tree whose root is $(n,\ell)$. \cbend

\begin{proposition}[Asymptotic behavior of the variance]\label{vartree}
When the size $N$ of the original tree goes to infinity, the variance of the size of the
sampled tree is such that
\begin{enumerate}
\item If the random variable $G$ is not deterministic and $\E(G^2)<+\infty$, then
\begin{equation}\label{rho21}
\rho_2^{(1)}(\lambda)\stackrel{\mathrm{def}}{=}  \lim_{N\to \infty}\frac{\Var(R_N)}{\E(T_N)^2} = \frac{\Var(G)}{m^2-m} \rho(\lambda)^2
\end{equation}
where $\rho(\lambda)$ is defined by Equation~\eqref{defrholambda}.
\item If $G\equiv m$ almost surely, then
\begin{multline} \label{rho22}
\rho^{(2)}_2(\lambda)\stackrel{\mathrm{def}}{=}  \lim_{N\to \infty}\frac{\Var(R_N)}{\E(T_N)} \
\ =
\frac{m-1}{m}\sum_{n=1}^{+\infty}
\frac{1}{m^n}\left[
\E\left(e^{-\lambda T_n}\right)\left(1-\E\left(e^{-\lambda T_n}\right)\right) \right. \\
\left.  +2\sum_{k=0}^{n-1}\E\left(e^{-\lambda T_{n{-}k{-}1}} Z_{n-k} \E\left(e^{-\lambda T_{k}
}\right)^{Z_{n-k}-1}
\E\left(e^{-\lambda T_{k}}\left(1-e^{-\lambda T_{k}}\right)\right)\right)\right].
\end{multline}
where $T_n = (m^{n+1}-1)/(m-1)$.
\end{enumerate}
\end{proposition}
It is worth  noting that the case of a deterministic  offspring distribution is degenerate
in the sense  that the standard deviation of  the size of the tree discovered  by means of
the traceroute  algorithm does not  scale with  the size of  the tree (and  the discovered
tree).  The  coefficient of variation of  the random variables  $R_N$ tends to 0  when $N$
goes to infinity.

\begin{proof}
\cbstart Using Representation~\eqref{relfond} for the size of the sampled tree, one obtains by conditioning on the tree the relation 
$$
R_N-\E(R_N) =A_{N,1}+A_{N,2}+A_{N,3},
$$
where 
\begin{align*}
A_{N,1} &=  \sum_{n=0}^N\sum_{\ell=1}^{Z_{N-n}} \Delta^\ell_n,\\
A_{N,2} &= \sum_{n=0}^N (Z_{N-n} - \E(Z_{N-n})) \P\left({\cal N}(\T_{n})\not=0\right),\\
A_{N,3} &= \sum_{n=0}^N \sum_{\ell=1}^{Z_{N-n}} \left(1-\exp(-\lambda T^{N-n,\ell}_n)-\P(\cal{N}(\T_n)\neq 0)\right)
\end{align*}
with
$
\Delta^\ell_n =\ind{{\cal N}(\T_{n}^{N-n,\ell})\not=0}  - (1-\exp(-\lambda
T_n^{N-n,\ell})).
$
Note that if distinct nodes $(n,l)$ and $(n',l')$ cannot be compared with the relation
$''{<}''$ then, conditionally on the tree ${\cal T}$, the corresponding random variables
$\Delta^\ell_n$ and $\Delta^{\ell'}_{n'}$ are {\em centered} and {\em independent}. 
In addition, note that $ A_{N,1} = R_N-\E(R_N~|~\T)$. 

To study the variance of the random variable $R_N$, we separately consider the second
moments of the terms $A_{N,1}$, $A_{N,2}$ and $A_{N,3}$. Of course, the terms $A_{N,2}$ and $A_{N,3}$ are  non null if
and only if the variable $G$ is not deterministic.  

\cbend

\subsection*{The second moment of $A_{N,1}$}
\cbstart 
It is shown that the second moment of $A_{N,1}$ is of the order of $m^N$.
By using the independence in the
selection of nodes in the tree and the fact that the random variables $\Delta_n^\ell$ are 
centered conditionally on $\T$, we have  the identity 
\[
\E(A_{N,1}^2~|~\T) = \sum_{(n,\ell)\in \T_N}\Var(\Delta^\ell_n~|~\T) + 2\sum_{\substack{(n,\ell),    (n',\ell')\in{\cal T}_N \\(n',\ell')<(n,\ell)}} \E\left(\Delta^\ell_n\Delta^{\ell'}_{n'}~|~\T\right).
\]
Conditioning on the state of the tree, when $(n',\ell')<(n,\ell)$, one has the identity
\[
\E\left(\Delta^\ell_n\Delta^{\ell'}_{n'}\mid {\cal T}\right)=
\exp\left(-\lambda T^{N-n,\ell}_{n}\right)\left(
1-\exp\left(-\lambda T^{N-n',\ell'}_{n'}\right)\right).
\]
By symmetry, the above computations yield the following relation for the second moment $\E(A_{N,1}^2)$
\begin{align*}
U_N&\stackrel{\text{def.}}{=} \E(A_{N,1}^2)-\sum_{n=0}^N\E(Z_{N-n})\Var(\Delta_n^1)\\
&= 2\E\left(\sum_{n=0}^N  \E(Z_{N-n}) \exp\left(-\lambda T_{n}^{N-n,1}\right)\sum_{\substack{(N-n',l')\in{\cal T}_N\\ (N-n',l')<(N-n,1)}} 
\left(1-\exp\left(-\lambda T_{n'}^{N-n',\ell'}\right)\right)\right),
\end{align*}
where $\Var(\Delta_n)$ is the variance of the random variable $\ind{\mathcal{N}(\T_n)\neq 0}-\P(\mathcal{N}(\T_n)\neq 0)$.
\cbend

For two nodes of the tree such that $(N{-}n',l')<(N{-}n,1)$, Equation~\eqref{eq1st} gives
the relation 
\[
T_n\stackrel{\text{dist.}}{=}
\sum_{k=0}^{n-n'-1}\widetilde{Z}_{k}+ 
\sum_{\ell'=1}^{\widetilde{Z}_{n-n'}} {T}_{n'}^{N-n',\ell'},
\]
where $(\widetilde{Z}_{k}, k\geq 0)$ denotes another independent Galton-Watson process
independent of $(Z_n, n \geq 0)$ with the same offspring distribution. By using this relation, we have
\begin{multline*}
\E\left(\exp\left(-\lambda T_{n}^{N-n,1}\right)\sum_{\substack{(N-n',l')\in{\cal T}_N\\ (N-n',l')<(N-n,1)}}\left(1-\exp\left(-\lambda T_{n'}^{N-n',\ell'}\right)\right)\right) \\ = \E\left(\sum_{n'=0}^{n-1}    \sum_{\ell'=1}^{Z_{n-n'}} \exp\left(-\lambda T_{n}^{N-n,\ell}\right)\left(1-\exp\left(-\lambda T_{n'}^{N-n',\ell'}\right)\right)\right)\\
= \E\left(\sum_{n'=0}^{n-1} \exp\left(-\lambda\sum_{k=0}^{n-n'-1} Z_k\right)  \E(V_{n-n'})\right),
\end{multline*}
where
$$
V_{n-n'} =  \sum_{\ell'=1}^{Z_{n-n'}} \exp\left(-\lambda\sum_{\ell''=1}^{Z_{n-n'}} T_{n'}^{N-n',\ell''}\right)\left(1-\exp\left(-\lambda T_{n'}^{N-n',\ell'}\right)\right).
$$
By using the independence of the different trees $\mathcal{T}_{n'}^{N-n',\ell'}$ for $\ell =1, \ldots, Z_{n-n'}$, we have
\begin{multline*}
\E(V_{n-n'}\mid Z_0,\ldots, Z_{n-n'-1}) \\ = \E\left(Z_{n-n'}\left(\E\left(
e^{-\lambda T_{n'}}\right)\right)^{Z_{n-n'}-1}\mid Z_0,\ldots,
Z_{n-n'-1}\right)\E\left(  e^{-\lambda T_{n'}} \left(1-e^{-\lambda T_{n'}}\right)\right). 
\end{multline*}
It follows that by using the above  expression for $U_N$, one obtains
\begin{multline*}
U_N=2\sum_{n=0}^N  \E(Z_{N-n})\sum_{n'=0}^{n-1}\E\left(\exp\left(-\lambda \sum_{k=0}^{n-n'-1}{Z}_{k}\right) \right.\\   
\left.\rule{0mm}{7mm} Z_{n-n'} \E\left(\rule{0mm}{4mm}\exp\left(-\lambda T_{n'}\right)\right)^{Z_{n-n'}-1}
\E\left(\rule{0mm}{4mm}\exp\left(-\lambda T_{n'}\right)\left(1-\exp\left(-\lambda
T_{n'}\right)\right)\right) \right).
\end{multline*}
Dividing by $\E(T_N)$, we have
\begin{multline*}
\frac{U_N}{\E(T_N)}=\frac{2(m-1)}{m-1/m^n}\sum_{n=0}^N  \frac{1}{m^n}\sum_{n'=0}^{n-1}\E\left(
\exp\left(-\lambda \sum_{k=0}^{n-n'-1}{Z}_{k}\right) \right.\\
\left.\rule{0mm}{7mm} Z_{n-n'}\E\left(\rule{0mm}{4mm}\exp\left(-\lambda T_{n'}\right)\right)^{Z_{n-n'}-1}
\E\left(\rule{0mm}{4mm}\exp\left(-\lambda T_{n'}\right)\left(1-\exp\left(-\lambda T_{n'}\right)\right)\right)\right).
\end{multline*}
By letting $N$ go to infinity, we finally obtain the relation for the second moment of the random variable $A_{N,1}$
\begin{multline}\label{A1}
\lim_{N\to \infty}\frac{\E(A_{N,1}^2)}{\E(T_N)} = 
\frac{m-1}{m}\sum_{n=1}^{+\infty}
\frac{1}{m^n}\left[
\E\left(e^{-\lambda T_n}\right)\left(1-\E\left(e^{-\lambda T_n}\right)\right) \right. \\
\left.  +2\sum_{k=0}^{n-1}\E\left(e^{-\lambda T_{n{-}k{-}1}} Z_{n-k} \E\left(e^{-\lambda T_{k}}\right)^{Z_{n-k}-1}
\E\left(e^{-\lambda T_{k}}\left(1-e^{-\lambda T_{k}}\right)\right)\right)\right].
\end{multline}

\subsection*{The second moment of $A_{N,2}$}
We have
\begin{multline*}
\frac{A_{N,2}}{m^{N}}  = \sum_{n=0}^{N} \frac{\left(Z_{N-n} - \E(Z_{N-n})\right)}{m^{N-n}}\frac{(1-\E(e^{-\lambda T_n}))}{m^n}
\\= \sum_{n=0}^{N} \left(\frac{Z_{N-n}}{m^{N-n}} - 1\right) \frac{(1-\E(e^{-\lambda T_n}))}{m^n}.
\end{multline*}
If $\|H\|_2=\sqrt{\E(H^2)}$ for some random variable $H$, then  we have
\[
\left\|\frac{A_{N,2}}{m^{N}}-(W - 1)\sum_{n=0}^{N} \frac{(1-\E(e^{-\lambda T_n}))}{m^n}\right\|_2 \leq \sum_{n=0}^{N} \left\|W-\frac{Z_{N-n}}{m^{N-n}}\right\|_2\frac{(1-\E(e^{-\lambda T_n}))}{m^n},
\]
where $W$ is defined by Equation~\eqref{defW}. \cbstart Athreya and Ney~\cite[Theorem~2, page~9]{Athreya:04} gives that the sequence
$\left(\left\|W-{Z_{n}}/{m^{n}}\right\|_2\right)$ converges to $0$. This implies
\begin{equation}\label{A2}
\lim_{N\to+\infty} \frac{\E(A_{N,2}^2)}{\E(T_N)^2}=\frac{(m-1)\Var(G)}{m^3} 
\left(\sum_{n=0}^{+\infty} \frac{(1-\E(e^{-\lambda T_n}))}{m^n}\right)^2.
\end{equation}
since
$\E\left((1-W)^2\right)={\Var(G)}/{m(m-1)}$. 

\subsection{Second moment of $A_{N,3}$} 
Clearly
$$
\|A_{N,3}\|_2 \leq \sum_{n=0}^N\left\|  \sum_{\ell=1}^{Z_{N-n}} \exp(-\lambda T^{N-n,\ell}_n) - \E(\exp(-\lambda T_n))\right\|_2 ,
$$
and since conditionally on $Z_{N-n}$, the random variables $\exp(-\lambda T^{N-n,\ell}_n)$
for $\ell = 1, \ldots, Z_{N-n}$ are independent and identically distributed with mean
$\E(\exp(- \lambda T_n))$,  we then have 
\begin{align*}
\left\|\sum_{\ell=1}^{Z_{N-n}} \exp(-\lambda T^{N-n,\ell}_n) - \E(\exp(-\lambda T_n))  \right\|_2^2 &= \E(Z_{N-n})\Var(\exp(-\lambda T_n))\\ & \leq m^{(N-n)}.
\end{align*}
It follows that 
$$
\|A_{N,3}\|_2 \leq \frac{m^{(N+1)/2}}{\sqrt{m}-1}.
$$
and then
\begin{equation}
\label{limA3N}
\limsup_{N\to\infty}\frac{\E(A_{N,3}^2)}{m^N} \leq \frac{m}{(\sqrt{m}-1)^2}.
\end{equation}

Since $R_N-\E(R_N)=A_{N,1}+A_{N,2}+A_{N,3}$, 
Relations~\eqref{A1}, \eqref{A2} and \eqref{limA3N} then imply  that  
\begin{enumerate}
\item When $G$ is non-deterministic, the expression $A_{N,2}$ dominates in $R_N-\E(R_N)$
so that $\Var(R_N)/\E(T_N)^2$ is converging to   the right hand side of Equation~\eqref{A2}. 
\item If $G\equiv m$, the term $A_{N,3}$ vanishes so that  $\Var(R_N)/\E(T_N)$ is converging to
  the right hand side of Equation~\eqref{A1}. 
\end{enumerate}
Equations~\eqref{rho21} and \eqref{rho22} are established.
\end{proof}

\cbend

As  for the  first  moment of  $R_N$, we  turn  now to  the  analysis of  the behavior  of
of the second order characteristics defined by Equations~\eqref{rho21} and \eqref{rho22}
when $\lambda$ is in the neighborhood of $0$. For the non deterministic case, we have from Proposition~\ref{rho1}
\[
\lim_{\lambda\to 0} \frac{\rho^{(1)}_2(\lambda)}{(\lambda\log_m(1/\lambda))^2} =\frac{\Var(G)}{(m^2-m)}.
\]
In  Proposition~\ref{vartree},  the expression  of  $\rho^{(2)}_2(\lambda)$  is defined  a
priori  only for  a  deterministic offspring  distribution,  but can  be  extended to  any  offspring   distribution  by   using  the   right  hand   side  of
Equation~\eqref{rho22}.     In    the    following,    we   study    the    behavior    of
$\rho_2^{(2)}(\lambda)$ for an arbitrary offspring distribution.

\begin{lemma}[Asymptotic Behavior of $\lambda{\to}\rho^{(2)}_2(\lambda)$ at $0$]\label{Rho22}
Provided that the random variable $G$ has a finite second moment, the function $\rho^{(2)}_2(\lambda)$ defined by
Equation~\eqref{rho22} is such that  
\begin{equation}\label{rate2}
\lim_{\lambda\searrow 0}\frac{\rho^{(2)}_2(\lambda)}{\lambda(\log_m\lambda)^2}=1.
\end{equation}
\end{lemma}

\begin{proof}
Define
$$
f_a(\lambda)\stackrel{\text{def.}}{=} \sum_{n=1}^{+\infty} \frac{m-1}{m^n}\left(\E\left(e^{-\lambda T_n}\right)\left(1-\E\left(e^{-\lambda T_n}\right)\right)\right)
$$
and 
\begin{multline*}
f_b(\lambda)\stackrel{\text{def.}}{=} \\ \sum_{n=1}^{+\infty} \frac{m-1}{m^n} \sum_{k=0}^{n-1}\E\left(e^{-\lambda T_{n{-}k{-}1}} Z_{n-k}\E\left(e^{-\lambda T_{k}}\right)^{Z_{n-k}-1}\right)\E\left(e^{-\lambda T_{k}}\left(1-e^{-\lambda T_{k}}\right)\right).
\end{multline*}
Equation~\eqref{rho22} gives that
${m}\rho^{(2)}_2(\lambda)=2f_b(\lambda)+f_a(\lambda)$. 

\medskip

\paragraph{\em Asymptotic behavior of $f_a$}
With similar arguments as in the proof of
Theorem~\ref{ThFirst} the asymptotic behavior of  $f_a(\lambda)$ when $\lambda$ goes to
$0$  is equivalent to the  asymptotic behavior of
\[
\sum_{n=1}^{+\infty}
\frac{m-1}{m^n}\left(\E\left(\exp\left(-\lambda \frac{W m^{n+1}}{m-1} \right)\right)\left(1-\E\left(\exp\left(-\lambda \frac{W m^{n+1}}{m-1} \right)\right)\right)\right).
\]
If $W_1$ and $W_2$ are two independent random variables with the same distribution as $W$,
the above series can be rewritten as
\begin{multline*}
\sum_{n=1}^{+\infty}
\frac{m-1}{m^n}
\E\left(\exp\left(-\lambda \frac{W_1 m^{n+1}}{m-1} \right)
\left(1-\E\left(\exp\left(-\lambda \frac{W_2 m^{n+1}}{m-1} \right)\right)\right)\right)
\\=
(m-1)\sum_{n=1}^{+\infty}
\left(\frac{1}{m^n} \E(h(\lambda(W_1+W_2)m^n/(m-1)))-\E(h(\lambda W_1m^n/(m-1)))\right),
\end{multline*}
with $h(u)=1-e^{-u}$. Consequently,
\begin{equation}\label{G1}
\lim_{\lambda\to 0} \frac{f_a(\lambda)}{-\lambda\log_m \lambda}=m
\end{equation}
by Proposition~\ref{asympprop}.

\medskip
\paragraph{\em Asymptotic behavior of $f_b$} Let us fix some $\varepsilon >0$ and assume that $\lambda <\varepsilon$.  The function $f_b(\lambda)$ can be rewritten as
\begin{equation}
\label{decompofb}
f_b(\lambda) = \sum_{n=1}^{\lfloor \log_m(\varepsilon/\lambda)\rfloor} \frac{m-1}{m^n} S(n;\lambda)  +  \sum_{n=\lfloor \log_m(\varepsilon/\lambda)\rfloor+1}^\infty  \frac{m-1}{m^n} S(n;\lambda) 
\end{equation}
where
$$
S(n;\lambda) =  \sum_{k=0}^{n-1}\E\left(e^{-\lambda T_{n{-}k{-}1}} Z_{n-k}\E\left(e^{-\lambda T_{k}}\right)^{Z_{n-k}-1}\right)\E\left(e^{-\lambda T_{k}}\left(1-e^{-\lambda T_{k}}\right)\right)
$$
Since for $x\geq 0$, $e^{-x}(1-e^{-x}) \leq x$, we easily deduce that for all $n \geq 1$
$$
S(n;\lambda) \leq  \sum_{k=0}^{n-1}\E\left( Z_{n-k}\right) \E\left(\lambda T_{k}\right) \leq \frac{n \lambda m^{n+1}}{m-1}
$$
and then
$$
\sum_{n=1}^{\lfloor \log_m(\varepsilon/\lambda)\rfloor} \frac{m-1}{m^n} S(n;\lambda) \leq \frac{m}{2}\lambda \log_m(\varepsilon/\lambda) (\log_m(\varepsilon/\lambda)+1).
$$

The second term in the right hand side of Equation~\eqref{decompofb} can be written as
\begin{multline}
\label{termtech2}
\sum_{n=\lfloor \log_m(\varepsilon/\lambda)\rfloor+1}^\infty  \frac{m-1}{m^n}      S( \lfloor \log_m(\varepsilon/\lambda)\rfloor +1;\lambda)  
 \\ 
+ \sum_{n=\lfloor \log_m(\varepsilon/\lambda)\rfloor+1}^\infty  \frac{m-1}{m^n}  \left(S( n;\lambda)  - S( \lfloor \log_m(\varepsilon/\lambda)\rfloor +1;\lambda)  \right).
\end{multline}
By using the fact that for $x>0$ and $\alpha>0$, $xe^{-\alpha x}\leq 1/\alpha$, we get that 
$$
 S( \lfloor \log_m(\varepsilon/\lambda)\rfloor +1;\lambda)  \leq \frac{1}{\E(e^{-\lambda T_k})}  \sum_{k=0}^{ \lfloor \log_m(\varepsilon/\lambda)\rfloor} \frac{\lambda\E(T_k)}{-\log\E(e^{-\lambda T_k})}. 
$$
The relation $\E\left(\exp\left(-\lambda T_k\right)\right) \geq 
\exp\left(-\lambda \E(T_k)\right)
\geq \exp(- m\varepsilon/(m-1))$  holds by Jensen's Inequality under the condition that $k \leq \lfloor
\log_m(\varepsilon/\lambda)\rfloor $. In addition,  
$$
\E(T_n^2) \leq \left(\sum_{i=0}^n \sqrt{\E(Z_i^2)}\right)^2 \leq \frac{m^{2n}}{(m-1)^2} \left(\frac{\sigma^2}{m-1}+1\right) ,
$$
where $\sigma^2$ is the variance of the random variable $G$, so that for $k \leq  \lfloor \log_m(\varepsilon/\lambda)\rfloor $
\begin{equation}
\label{rapportmoment}
\frac{\lambda\E(T_k^2)}{\E(T_k)} \leq \frac{\lambda m^k m^k}{(m-1)(m m^k-1)} \left(\frac{\sigma^2}{m-1}+1\right)  \leq \frac{\varepsilon}{(m-1)^2} \left(\frac{\sigma^2}{m-1}+1\right). 
\end{equation}
Since for $x \geq 0$, $e^{-x} \leq 1-x+{x^2}/{2}$, we have
$$
\E(e^{-\lambda T_k}) \leq 1 -\lambda\E(T_k) +\frac{\E\left((\lambda T_k)^2\right)}{2} \leq 1-\lambda \E(T_k) \left(1-  \frac{\varepsilon}{(m-1)^2} \left(\frac{\sigma^2}{m-1}+1\right) \right).
$$
and then, for $k \leq  \lfloor \log_m(\varepsilon/\lambda)\rfloor $, 
\begin{multline}\label{eqaux}
\frac{\lambda\E(T_k)}{-\log\E(e^{-\lambda T_k})} \leq
\frac{\lambda\E(T_k)}{1- \E(e^{-\lambda T_k}) }\\ \leq
\left(1-  \frac{\varepsilon}{(m-1)^2} \left(\frac{\sigma^2}{m-1}+1\right)\right)^{-1}
 \stackrel{\mathrm{def.}}{=}\kappa(\varepsilon)
\end{multline}
as long as 
$$
\varepsilon< (m-1)^2\left/\left(\frac{\sigma^2}{m-1}+1\right)\right. \stackrel{\mathrm{def.}}{=}\varepsilon_1.
$$

It follows that for $\varepsilon<\varepsilon_1$,
$$
S( \lfloor \log_m(\varepsilon/\lambda)\rfloor +1;\lambda)  \leq  (1+\log_m(\varepsilon/\lambda))  e^{m\varepsilon/(m-1)}\kappa(\varepsilon).
$$
and therefore,
$$
\sum_{n=\lfloor \log_m(\varepsilon/\lambda)\rfloor+1}^\infty  \frac{m-1}{m^n}      S( \lfloor \log_m(\varepsilon/\lambda)\rfloor +1;\lambda)  \leq \frac{\lambda}{\varepsilon}  \log_m(\varepsilon/\lambda)  e^{m\varepsilon/(m-1)}\kappa(\varepsilon),
$$
which is $o(\lambda(\log\lambda)^2)$ when $\lambda \to 0$. In addition, the second term in the right hand side of Equation~\eqref{termtech2} can be rewritten as
$$
  \sum_{k= \lfloor \log_m(\varepsilon/\lambda)\rfloor +1}^\infty \frac{m-1}{m^k} \sum_{n=1}^\infty \frac{1}{m^n} \E\left(e^{-\lambda T_{n-1}} Z_n \E(e^{-\lambda T_k})^{Z_n-1}\right)  \E\left(e^{-\lambda T_{k}}\left(1-e^{-\lambda T_{k}}\right)\right).
$$
We first note that
\begin{multline*}
\sum_{n=1}^\infty \frac{1}{m^n} \E\left(e^{-\lambda T_{n-1}} Z_n \E(e^{-\lambda T_k})^{Z_n-1}\right) =\\
\sum_{n=1}^{\lfloor \log_m(\varepsilon/\lambda)\rfloor} \frac{1}{m^n} \E\left(e^{-\lambda T_{n-1}} Z_n \E(e^{-\lambda T_k})^{Z_n-1}\right) \\ 
+ \sum_{n = \lfloor \log_m(\varepsilon/\lambda)\rfloor+1}^\infty \frac{1}{m^n} \E\left(e^{-\lambda T_{n-1}} Z_n \E(e^{-\lambda T_k})^{Z_n-1}\right).
\end{multline*}
The first term in the right hand side of the above equation is less than or equal to the
quantity $\log_m(\varepsilon/\lambda)$ since $\E(Z_n)=m^n$. The second term can be upper bounded as
\begin{multline*}
\sum_{n = \lfloor \log_m(\varepsilon/\lambda)\rfloor+1}^\infty \frac{1}{m^n} \E\left(e^{-\lambda T_{n-1}} Z_n \E(e^{-\lambda T_{k}})^{Z_n-1}\right) \\ \leq \sum_{n=\lfloor \log_m(\varepsilon/\lambda)\rfloor+1}^\infty \frac{1}{m^n}\E\left(Z_n  e^{-\lambda Z_n} \right) 
\leq \frac{1}{(m-1)\varepsilon},
\end{multline*}
where we have used the fact that $T_k \geq 1$ for all $k \geq 0$ and $xe^{-\lambda x}\leq 1/(e\lambda)$ for all $x>0$. It follows that the second term in the right hand side of Equation~\eqref{termtech2} is upper bounded by the quantity 
$$
\frac{\lambda}{\varepsilon}\left(\log_m(\varepsilon/\lambda)+ \frac{1}{(m-1)\varepsilon}\right),
$$
which is $o(\lambda(\log_m\lambda)^2)$ when $\lambda \to 0$. 

By using the above inequalities, we come up with the conclusion that for every $\varepsilon >0$,
\begin{equation}
\label{limsupfb}
\limsup_{\lambda \to 0} \frac{f_b(\lambda)}{\lambda(\log_m\lambda)^2} \leq \frac{m}{2}.
\end{equation}

For establishing a lower bound for $f_b(\lambda)$, we introduce the size-biased Galton-Watson branching process.  The sequence of random variables $(Z_n/m^n)$ being a positive  martingale, it induces a probability distribution $\widetilde{\P}$ such that, for any $n\geq 1$  and any random variable $Y$ measurable with respect to the random variables $Z_1, \ldots, Z_n$, 
 \[
 \int Y d\widetilde{\P}=\E\left(Y\frac{Z_n}{m^n}\right).
 \]
 It is known, see Lyons and Peres~\cite{Lyons:07}, that under the probability
 $\widetilde{\P}$, the sequence $(Z_n)$ as the same distribution as  a branching process with
 immigration $(\widetilde{Z}_n)$ where the number of children has the same distribution as
 $G$ and the number of new immigrants is distributed as $\widetilde{G}$ such that
 $\P(\widetilde{G}=n)=n\P(G=n)/m$. If $\widetilde{Z}_0=1$, it is easy to check that
 \[
 \widetilde{\E}\left(\widetilde{Z}_n\right)=m^n+\frac{m^n-1}{m(m-1)}\E(G^2).
 \]
 If $\widetilde{T}_n=\widetilde{Z}_0+\widetilde{Z}_1+\cdots+\widetilde{Z}_n$, we have by Jensen inequality
\begin{align*}
\E\left(\rule{0mm}{5mm}e^{-\lambda T_{n{-}k{-}1}}\right.&\left. \frac{Z_{n-k}}{m^{n-k}} \E\left(e^{-\lambda T_{k}}\right)^{Z_{n-k}-1}\right) =\widetilde{\E}\left(e^{-\lambda \widetilde{T}_{n{-}k{-}1}} \E\left(e^{-\lambda T_{k}}\right)^{\widetilde{Z}_{n-k}-1}\right) \\
&\geq  \widetilde{\E}\left(e^{-\lambda \widetilde{T}_{n{-}k}} \E\left(e^{-\lambda T_{k}}\right)^{\widetilde{T}_{n-k}}\right) \\
&\geq \exp\left(-\lambda(1+\E(T_k)) \widetilde{\E}( \widetilde{T}_{n{-}k})\right)\\
&\geq \exp\left(-\frac{\lambda m}{m-1} \left(m^{n-k}+\frac{m^{n+1}}{(m-1)}\right)\left(1+\frac{g_2}{m}\right)\right)
\end{align*}
since 
$$
\widetilde{\E}( \widetilde{T}_{n}) \leq \frac{m^{n+1}}{m-1}\left(1+\frac{g_2}{m}  \right),
$$
where $g_2 = \E(G^2)$. In addition, by using the fact that $e^{-x}(1-e^{-x}) \geq x-2x^2$
holds for $x>0$, we have
\begin{multline*}
 \sum_{n=1}^{\lfloor \log_m(\varepsilon/\lambda)\rfloor} \frac{m-1}{m^n} S(n;\lambda)  
\geq  \sum_{n=1}^{\lfloor \log_m(\varepsilon/\lambda)\rfloor} (m-1) \\\times
\sum_{k=0}^{n-1} \frac{1}{m^k} 
\exp\left(-\frac{\lambda m}{m-1} \left(m^{n-k}+\frac{m^{n+1}}{(m-1)}\right)
\left(1+\frac{g_2}{m}\right)\right)
\E\left(\lambda T_k-2(\lambda T_k)^2\right) \\
\geq \exp\left(-\frac{\varepsilon m}{(m-1)}
  \left(1+\frac{m}{(m-1)}\right)\left(1+\frac{g_2}{m}\right)\right)\\ \sum_{n=1}^{\lfloor
  \log_m(\varepsilon/\lambda)\rfloor} (m-1) \sum_{k=0}^{n-1}\frac{\lambda\E(T_k)}{m^k}
\left(1-\frac{\lambda\E(T_k^2)}{\E(T_k)}\right). 
\end{multline*}
By using Inequality~\eqref{rapportmoment} and Definition~\eqref{eqaux}, we have, for
$\eps<\eps_1$, 
\begin{multline*}
\sum_{n=1}^{\lfloor \log_m(\varepsilon/\lambda)\rfloor} (m-1) \sum_{k=0}^{n-1}\frac{\lambda\E(T_k)}{m^k} \left(1-\frac{\lambda\E(T_k^2)}{\E(T_k)}\right)  \geq \frac{\lambda}{\kappa(\eps)}\sum_{n=1}^{\lfloor \log_m(\varepsilon/\lambda)\rfloor} \sum_{k=0}^{n-1}\frac{m^{k+1}-1}{m^k}
\end{multline*}
Since
\begin{multline*}
\sum_{n=1}^{\lfloor \log_m(\varepsilon/\lambda)\rfloor} \sum_{k=0}^{n-1}\frac{m^{k+1}-1}{m^k} = \frac{m}{2}\lfloor \log_m(\varepsilon/\lambda)\rfloor(\lfloor \log_m(\varepsilon/\lambda)\rfloor+1)+\frac{m\lfloor \log_m(\varepsilon/\lambda)\rfloor}{m-1} \\ -\frac{m}{(m-1)^2}\left(\frac{1}{m^{\lfloor \log_m(\varepsilon/\lambda)\rfloor}}-1\right)
\end{multline*}
and since we already know that the second term in the right hand side of Equation~\eqref{decompofb} is $o(\lambda(\log_m(\lambda))^2$ when $\lambda\to 0$, we then deduce that for all $\varepsilon \in (0,\varepsilon_1)$
$$
\liminf_{\lambda \to 0} \frac{f_b(\lambda)}{\lambda(\log_m\lambda)^2} \geq
\frac{m}{2\kappa(\eps)} \exp\left(-\frac{\varepsilon m}{(m-1)} \left(1+\frac{m}{(m-1)}\right)\left(1+\frac{g_2}{m}\right)\right)
$$
and hence,
\begin{equation}
\label{liminffb}
\liminf_{\lambda \to 0} \frac{f_b(\lambda)}{\lambda(\log_m\lambda)^2} \geq \frac{m}{2}.
\end{equation}
Combining Equations~\eqref{G1}, \eqref{limsupfb} and \eqref{liminffb}, Equation~\eqref{rate2} follows.
 \end{proof}

\begin{proposition}
The functions $\rho^{(1)}_2(\lambda)$ and $\rho^{(2)}_2(\lambda)$ are such that
\begin{eqnarray}
\lim_{\lambda \to 0}\frac{\rho^{(1)}_2(\lambda)}{(\lambda\log_m\lambda)^2} &=& \frac{1}{m^2-m}\Var(G),\\ 
\lim_{\lambda \to 0} \frac{\rho^{(2)}_2(\lambda)}{\lambda(\log_m\lambda)^2} &=& 1.
\end{eqnarray}
where $G$ is the random variable  describing the offspring of a node.
\end{proposition}

From Theorem~\ref{rho1},  we observe  that the size  of the sampled tree  scales with the
size  of  the  original  tree.  The   same  phenomenon  is  true  for  the  squared coefficient of variation  of the size of the sampled tree if an only if the offspring distribution is not deterministic as shown by Proposition~\ref{vartree}. In the case of a deterministic offspring distribution, when $\lambda \to 0$,  the squared coefficient of variation is approximately equal to $1/(\lambda \E(T_N))$ for large $N$. The quantity $\lambda\E(T_N)$ is precisely the mean number of selected points. This indicates that the distribution of the random variable $R_N$ is concentrated around its mean value.  There is almost no randomness in the discovered tree.

\section{The Depth Biased Model}
\label{biased}
In this section,  it is assumed that conditionally  on the tree, for $n\geq 0$,  a node at
depth   $n$    is   chosen    with   probability   $(1-\exp(-(\alpha/m)^n))$    for   some
$\alpha\in[0,1)$. The mean number  of selected nodes at depth $n$ in  the tree is equal to
$m^n (1-\exp(-(\alpha/m)^n))\sim \alpha^n$ and the total number of selected nodes in the
whole tree $\mathcal{N}(\mathcal{T})$ is such that
$$
\frac{1}{1-\alpha} -\frac{1}{2(1-\alpha^2/m)} \leq \E(\mathcal{N}(\mathcal{T})) = \sum_{n=0}^\infty m^n (1-e^{-(\alpha/m)^n})\leq \frac{1}{1-\alpha},
$$  in  particular  the  mean  number of  selected  nodes  $\mathcal{N}(\mathcal{T})  \sim
1/(1-\alpha)$ when  $\alpha \to 1$.  The behavior of  the size $R(\alpha)$ of  the sampled
tree is used to estimate the speed of the exploration process, when the number of selected
nodes becomes large. We first give the expression of the mean value $\E(R(\alpha))$ of the
size of the sampled tree.

\begin{lemma}
The mean value of the size of the sampled tree in the depth biased model is given by
\begin{equation}
\label{relfond1}
\E(R(\alpha)) = \sum_{n=0}^\infty m^n \left(1-\E\left(\exp\left(-\left(\frac{\alpha}{m}\right)^n\sum_{i=0}^\infty \alpha^i \frac{Z_i}{m^i}\right)\right)\right).
\end{equation}
\end{lemma}

\begin{proof}
As in the previous section, for $n\geq 0$ and $1\leq \ell\leq Z_n$, the symbol  ${\cal T}^{n,\ell}$
denotes the sub-tree of ${\cal T}$ whose root is $(n,\ell)$. The node $(n,\ell)$  is in the sampled tree if ${\cal N}({\cal T}^{n,\ell})\neq 0$. 
\cbstart Since nodes at a given depth are selected independently one of  each other, we have
$$ 
\P\left({\cal N}({\cal T}^{n,\ell})\not=0 \mid {\cal T}, (n,\ell) \in {\cal T}\right) =1 - \exp\left(-\left(\frac{\alpha}{m}\right)^n\sum_{i=0}^\infty \alpha^i \frac{Z^{(n,\ell)}_i}{m^i}\right),
$$
where $Z_i^{(n,\ell)}$ is the number of descendants of $(n,\ell)$ at generation $i$ and
where we have used the fact that the sub-tree ${\cal T}^{n,\ell}$ has the same offspring
distribution as the original tree $\mathcal{T}$. Hence,  
$$
\P\left({\cal N}({\cal T}^{n,\ell})\not=0  \mid (n,\ell) \in {\cal T}\right)   = 1 - \E\left(\exp\left(-\left(\frac{\alpha}{m}\right)^n\sum_{i=0}^\infty \alpha^i \frac{Z_i}{m^i}\right)\right).
$$ 
\cbend
It follows that the size of the sampled tree given by 
\begin{equation} 
R(\alpha) =\sum_{n=0}^{+\infty} \sum_{\ell=1}^{Z_{n}}\ind{{\cal N}({\cal T}^{\ell,n})\not=0}
\end{equation}
and its mean value is, by using the independence in the selection of nodes, \cbstart
$$
\E(R(\alpha)) =  \sum_{n=0}^{+\infty} \E(Z_n)\P\left({\cal N}({\cal T}^{n,1})\not=0\right),
$$
Equation~\eqref{relfond1} follows.\cbend
\end{proof}

The growth rate of the exploration process is defined  by the ratio
\[
\frac{\E\left(R(\alpha)\right)}{\E({\cal N}({\cal T}))} =\frac{1}{\eta(\alpha)} \sum_{n=0}^{+\infty} (1-\alpha)m^n
\left(1-\E\left(\exp\left(-\left(\frac{\alpha}{m}\right)^n\sum_{i=0}^\infty \alpha^i \frac{Z_i}{m^i}\right)\right)\right),
\]
where $\eta(\alpha)=(1-\alpha)\E({\cal N}({\cal T})) \to 1$ when $\alpha \to 1$.

\begin{theorem}\label{oscil}
If $\E(G^2)<+\infty$, as $\alpha\nearrow 1$, the following limit relation holds 
\[
\lim_{\alpha \to 1}\frac{\E\left(R(\alpha)\right)}{\E({\cal N}({\cal T}))^2} =1.
\]
\end{theorem}

\begin{proof}
Let us first introduce the function
$$
H(\alpha) = \sum_{n=0}^{+\infty} (1-\alpha)m^n \left(1-\E\left(\exp\left(-\left(\frac{\alpha}{m}\right)^n\frac{W}{1-\alpha}\right)\right)\right),
$$
where $W$ is defined by Equation~\eqref{defW}. We have
\begin{eqnarray}
\left|H(\alpha) - \eta(\alpha)\frac{\E\left(R(\alpha)\right)}{\E({\cal N}({\cal T}))}
\right| &\leq& (1-\alpha) \sum_{n=0}^\infty \alpha^n \left|\sum_{i=0}^\infty \alpha^i
\E\left(\frac{Z_i}{m^i} - W\right) \right|\label{eq0}\\  
& \leq &\sum_{i=0}^\infty \alpha^i \E\left(\left| \frac{Z_i}{m^i} - W  \right|\right)
\label{eqtech} \\
&\leq& \mathrm{Var}(W)  \sum_{i=0}^\infty \left(\frac{\alpha}{\sqrt{m}}\right)^i =\frac{\mathrm{Var}(W) }{1-\frac{\alpha}{\sqrt{m}}},\notag
\end{eqnarray}
where we have used Inequality~\eqref{CS} in the last step.

Let us define the family of non-negative random variables $\mathcal{H}_\alpha$, $0<\alpha<1$ by
$$
\mathcal{H}_\alpha = \sum_{n=0}^{+\infty} (1-\alpha)^2m^n \left(1-\exp\left(-\left(\frac{\alpha}{m}\right)^n\frac{W}{1-\alpha}\right)\right).
$$
We have $(1-\alpha)H(\alpha)= \E(\mathcal{H}_\alpha )$. 

Let us fix some $\varepsilon >0$. Since $W>0$ a.s., we can define the quantity 
$$n(W,\alpha)=\max\left(\left\lceil
\log_{m/\alpha}\left(\frac{W}{(\varepsilon(1-\alpha))}\right)\right\rceil, 0\right).$$
For $n\geq n(W,\alpha)$, 
$$
\left( \frac{\alpha}{m} \right)^n\frac{W}{1-\alpha} <\varepsilon.
$$
By using the fact that for $x\geq 0$, $1-e^{-x} \geq x - x^2/2$, we have
$$
\mathcal{H}_\alpha \geq  \sum_{n=n(W,\alpha)}^{+\infty} 
(1-\alpha)^2m^n
\left(1-\exp\left(-\left(\frac{\alpha}{m}\right)^n\frac{W}{1-\alpha}\right)\right)  \geq \alpha^{n(W,\alpha)} W(1-\varepsilon).
$$   Since  the   above  inequality   is  valid   for  all   $\varepsilon  >0$   and  
$\alpha^{n(W,\alpha)}$  converges   to  $1$  as   $\alpha\nearrow  1$,  it   follows  that
$\liminf_{\alpha\to 1}\mathcal{H}_\alpha \geq W$ a.s.

Since $1-e^{-x} \leq x$ for $x\geq 0$, we have
$$
\mathcal{H}_\alpha \leq   W \quad \mbox{a.s.}
$$
and then $\limsup_{\alpha\to 1}\mathcal{H}_\alpha \leq W$ a.s. Hence, $\limsup_{\alpha\to 1}\mathcal{H}_\alpha = W$ a.s. Since the family $(\mathcal{H}_\alpha)$ is non negative and bounded by $W$, which is integrable, we have
$$
\lim_{\alpha\to 1} \E(\mathcal{H}_\alpha )= \E(W)=1
$$
and the result follows by using Inequality~\eqref{eqtech}.
\end{proof}

When $\alpha <1$ the selected node are closed to the root and only a small fraction of the
whole is  discovered. When  $\alpha \nearrow 1$,  we can  select nodes deeper  in the  tree but
roughly only  one node is selected  in average at  each level. The above  result indicates
that the average  size of the discovered  tree grows as the square  of the average
number of selected nodes.

\bibliographystyle{amsplain}
\bibliography{main}

\end{document}